\documentclass[superscriptaddress, twocolumn, amsmath, amssymb, aps,
prapplied, letterdraft, notitlepage,longbibliography]{revtex4-1} 
\usepackage{graphicx,graphics,epsfig,subfigure,times,bm,bbm,amssymb,amsmath,amsfonts,amsthm,mathrsfs,MnSymbol}
\usepackage[matrix,frame,arrow]{xypic}
\usepackage[normalem]{ulem}
\usepackage{slashed}
\usepackage{color}
\usepackage[usenames,dvipsnames,svgnames,table]{xcolor}
\usepackage[english]{babel}
\definecolor{darkblue}{rgb}{0.0,0.0,0.3}
\usepackage[colorlinks=true,
            linkcolor=red,
            urlcolor= darkblue,
            citecolor=blue]{hyperref}

\newcommand{\bes} {\begin{subequations}}
\newcommand{\ees} {\end{subequations}}
\newcommand{\bea} {\begin{eqnarray}}
\newcommand{\eea} {\end{eqnarray}}

\def\>{\rangle}
\def\<{\langle}

\begin{document}

\title{Spectral Transfer Tensor Method for Non-Markovian Noise Characterization}

\author{Yu-Qin Chen}
\affiliation{Tencent Quantum Laboratory, Tencent, Shenzhen, Guangdong, China, 518057}
\author{Yi-Cong Zheng}  
\affiliation{Tencent Quantum Laboratory, Tencent, Shenzhen, Guangdong, China, 518057}
\author{Shengyu Zhang}
\affiliation{Tencent Quantum Laboratory, Tencent, Shenzhen, Guangdong, China, 518057}
\author{Chang-Yu Hsieh}
\email{kimhsieh@tencent.com}
\affiliation{Tencent Quantum Laboratory, Tencent, Shenzhen, Guangdong, China, 518057}

\begin{abstract}
With continuing improvements on the quality of fabricated quantum devices, it becomes increasingly crucial to analyze noisy quantum process in greater details such as characterizing the non-Markovianity in a quantitative manner. In this work, we propose an experimental  protocol, termed Spectral Transfer Tensor Maps  (SpecTTM), to accurately predict the RHP  non-Markovian measure of any Pauli channels without state-preparation and measurement (SPAM) errors. In fact, for Pauli channels, SpecTTM even allows the reconstruction of highly-precised noise power spectrum for qubits. At last, we also discuss how SpecTTM can be useful to approximately characterize non-Markovianity of non-Pauli channels via Pauli twirling in an optimal basis. 

\end{abstract}

\maketitle

\section{Introduction}

The past decade has witnessed a rapid development of quantum technologies, especially with the advent of the noisy intermediate-scale quantum (NISQ) era~\cite{preskill_qm_18}.  These achievements are firmly built upon our continually improving capability to fabricate, control, and benchmark quantum devices with unprecedented precision.  Characterization protocols like quantum state and process tomography, randomized benchmarking (RB), Hamiltonian learning, and quantum noise spectroscopy have all proven to be instrumental in the diagnosis of faulty quantum devices and, eventually, help us build higher-quality devices. Recently, increasing attention has been turned toward more precise noise characterizations and control, such as detecting and suppressing non-Markovian noises, which bears implications to other characterization protocols such as RB that assumes Markvoian noises. Despite these recent efforts, there is still no SPAM error-free method to efficiently quantify strength of non-Markovian noises in a quantum device.  In this work, we propose such a method by combining recently proposed transfer tensor method (TTM) and the spectral quantum process tomography (SpecQPT).

Quantum devices are often operated in a noisy environment. If the noise is Markovian, it is straightforward to predict the adverse effects and manipulate quantum states as desired.  When non-Markovian memory plays a critical role, the associated quantum dynamics becomes significantly harder to analyze and control as the dynamical evolution is intimately affected by the past trajectory. This challenge could even create non-trivial roadblocks for building a fault tolerant quantum computer. Hence, as we attempt to increase the scale of a quantum device, such as the circuit depth and qubit counts of a quantum circuit,  it is desirable to quantify and suppress non-Markovian noises.

Currently, there are many competing proposals for non-Markovianity measures due to different perspectives and motivations.  This is an active research area, and no unanimous consensus on the most useful measure (for developing quantum technology) has been reached. Among them, the RHP measure is inspired by the violation of CP-divisibility for non-Markovian dynamical maps.  While RHP is a mathematically rigorous approach, it is not straightforward to deduce in experiments. The challenge is further aggravated by the need to remove SPAM errors in order to achieve high-precision measurements for a quantum hardware.  
	
Quantum process tomography (QPT) \cite{chuang_prescription_1997,poyatos_complete_1997,nielsen_chuang_book00} is among the most widely used tool for experimentally characterizing quantum dynamics and has been used in the context of quantum technology, such as determining quantum gate fidelities \cite{bialczak_quantum_2010,yamamoto_quantum_2010,rodionov_dnrey_prb2014} and investigating environment-induced errors\cite{joelpnas11,howard_quantum_2006}. However, earlier theoretical and experimental efforts \cite{kofman_two-qubit_2009,howard_quantum_2006} were largely confined to device characterization without taking non-Markovianity of noise into account.  A newly proposed method, transfer tensor map (TTM) \cite{PhysRevLett.112.110401}, offers a way to bridge  experimentally deduced QPT data with the theory of time-nonlocal quantum master equation (TNQME) \cite{breuer_theory_2007}, which is valid for general open quantum dynamics including non-Markovian ones.  Built upon the theorey-experiment connection under the TTM framework, Ref.~\onlinecite{chen2020non} further proposes a suite of experimental protocols to more easily (1) witness the non-Markovianity of a quantum process, (2) reconstruct the noise power spectrum, and (3) estimate spatial and temporal correlations of non-local noise in a quantum device.

In this work, we replace the standard QPT experiments considered in Ref.~\onlinecite{chen2020non} with the newly proposed spectral QPT (SpecQPT) \cite{helsen2019spectral}, which gives SPAM error-free estimates of eigenvalues of quantum channels. The original SpecQPT is most suitable to characterize Markovian quantum channels. In the original proposal~\cite{helsen2019spectral}, non-Markovian noises can only be detected but not rigorously quantified. Combining SpecQPT with TTM, we derive a spectral form of TTM, dubbed SpecTTM, that allows us to quantify degrees of non-Markovianity for any Pauli channels.  With limited efforts, one can further extract the noise power spectrum for these channels.  For more general quantum processes beyond Pauli channels, we apply Pauli twirling in an optimal basis (as defined in a later section) to estimate the degree of non-Markovianity. Without loss of generality, we will describe the newly proposed SpecTTM in the context of characterizing qubits in a circuit model.

The remainder of this work is organized as follows. Sec.~\ref{sec:ttm_spect} summarizes the derivation of spectral transfer tensor maps. Sec.~\ref{sec:noise} discusses the proposed SpecTTM noise spectroscopy. Sec.~\ref{sec:nopauli} discusses an extension of SpecTTM to non-Pauli channel.
Sec.~\ref{sec:simulation} describes the usage of
SpecTTM noise spectroscopy with examples based on theoretical models.
Section VI concludes.

\section{Spectral transfer tensor method for Pauli channels}\label{sec:ttm_spect}

The spectral transfer tensor map (SpecTTM) is an elegant combination of TTM and SpecQPT for characterizing Pauli channels.  We first recap these two methods before we present our contribution, SpecTTM, which draws inspirations from both methods to efficiently investigate non-Markovian dynamics without influences of SPAM errors. This high-precision analysis should provide essential information to further improve the quality of existing quantum devices.

\subsection{Transfer tensor maps}\label{sec:ttm}

We first review the theory of TTM as originally formulated in Ref.~\onlinecite{PhysRevLett.112.110401}.  The most general dynamical evolution of an open quantum system is given by
\begin{eqnarray}\label{eq:dynmapp}
    \rho_t & = & \left \langle \exp_+\left(-i\int^t_0 ds H(s)\right) \rho_0\rho_B \exp_-\left(i\int^t_0 ds H(s)\right) \right \rangle \nonumber \\
&=& \Lambda_t \rho_0,
\end{eqnarray}
where $\rho_B$ is the initial state of the environment, $H(t)$ is the total Hamiltonian of a time-dependent system comprising a system and its environment. The $\pm$ subscript denotes the (anti-)chronological time ordering of the time-evolution operator. The bracket $\langle \cdots \rangle $ denotes an average over the environmental degrees of freedom. $\Lambda_t$ is the dynamical map relating  the initial density matrix to the time-evolved reduced density matrix. The dynamical maps $\Lambda$ for a $d$-level quantum system are derived from an ensemble of QPTs obtained under $d^2$ different initial conditions. In an experiment, the QPTs are supposedly performed at equidistant time intervals, i.e. $t_k = k \delta t$, thus $\Lambda_k \equiv \Lambda_{t_k}$.   

If we assume a separable system-bath initial condition and a time independent Hamiltonian then an open system's dynamics can be succinctly cast in the form,
\begin{eqnarray}\label{eq:ttm2}
\rho_{t_n} = \sum_{m=1}^{n} T_m \rho_{t_{n-m}},
\end{eqnarray}
where the system's state at time $t_n$ is determined the history of its past evolution extending all the way back to $t_0$. In Eq.~(\ref{eq:ttm2}), TTMs $T_m$ are introduced to correlate two density matrices $\rho_{t_n}$ and $\rho_{t_{n-m}}$.  More specifically, TTMs are defined via,
\begin{eqnarray}\label{eq:ttm}
T_n \equiv \Lambda_n - \sum_{m=1}^{n-1} T_{n-m}\Lambda_m,
\end{eqnarray}
with $T_1 = \Lambda_1$. Time translational invariance of the dynamical process is implicitly assumed, as these maps are related by the time difference $m\delta t$ . 

For most practical cases, one can approximate exact quantum dynamics by truncating the TTM series to a finite number of terms, i.e.$\{T_1, \cdots, T_M\}$ in Eq.~(\ref{eq:ttm2}). This observation is the key that the TTM formalism could be useful in an actual experiment, as one only needs to perform $M$ QPTs to deduce the dynamical maps $\{\Lambda_1, \cdots, \Lambda_M\}$ and associated TTMs. Beyond $t_M$, all quantum dynamical information can be recursively predicted  with the help of TTMs via Eqs.~(\ref{eq:ttm2})-(\ref{eq:ttm}).

\subsection{Spectral quantum process tomography }\label{sec:tomography1}
   
Spectral quantum process tomography (SpecQPT)~ \cite{helsen2019spectral} is another newly developed tomographic technique that only provides partial information (eigenvalues of a quantum channel) on the dynamical processes under investigation. SPAM-resistant protocols must make certain trade-off between characterization details and efficiency. For instance, scalable method like RB may only output a scalar number to benchmark the gate quality.  On the other hand, methods like robust tomography and gate-set tomography characterize all aspects of a quantum gate at the expense of scalability. Aforementioned examples are extreme cases, and SpecQPT situates in the middle of the spectrum.  It gives SPAM error-free estimates on the spectral properties of a quantum gate without consuming as much resources as the standard QPTs, which are SPAM error-prone.  Furthermore, it does not require complex experimental protocols for implementations. We briefly summarize the idea of SpecQPT below.

Any CPTP map $\Lambda$ acting on density operators in a $d$-dimensional Hilbert space for an $n$-qubit system can be represented with a corresponding Pauli transfer matrix,
\begin{equation}   \label{eq:PTA}
S_{\mu \nu}=\operatorname{Tr}\left[P_{\mu} \Lambda\left(P_{\nu}\right)\right], \quad \mu, \nu=0, \ldots, N,
\end{equation}
where $P_{0}=I^{\otimes n},\quad P_{1}=I^{\otimes n-1} \otimes X, P_{2}=I^{\otimes n-1} \otimes Y$, ..., $P_N=Z^{\otimes n}$ are Pauli matrices with $N=d^2-1$ and $d=2^n$. If we are restricted to dealing with unital channels, i.e. $\Lambda(I)=I$, then the Pauli transfer matrix assumes a block-diagonal form,
\begin{equation}\label{eq:Smtx}
S=\left(\begin{array}{cc}1 & 0 \\ 0 & R\end{array}\right).
\end{equation}

The SpecQPT protocol relies on an important observation that eigenvalues of $R$ matrix in Eq.~(\ref{eq:Smtx}) may be deduced from a set of experimentally generated signal functions $\{g(1),g(2),...,g(K)\}$where $K\ge 2N-2$ in order to determine the eigenvalues accurately.
These signal functions are defined as
\begin{eqnarray}\label{eq:gk}
g(k)&=&\frac{1}{2^n}\sum_{\mu=1}^{N} \operatorname{Tr}\left[P_{\mu} \mathcal{N}_{\text {m }} \circ \Lambda^{k} \circ \mathcal{N}_{\text {p }}\left(P_{\mu}\right)\right], \nonumber \\
&=&\frac{1}{2^n}\operatorname{Tr}\left[R_{\text {meas }} R^{k} R_{\text {prep }}\right], \nonumber \\ 
&=&\frac{1}{2^n}\operatorname{Tr}\left[A_{\mathrm{SPAM}} D^{k}\right]=\frac{1}{2^n}\sum_{j=1}^{N} A_{j} \lambda_{j}^{k},
\end{eqnarray}
where $\mathcal{N}_{\text {p }}$, $\Lambda^k$, and $\mathcal{N}_{\text {m }}$ are quantum channels corresponding to the state-preparation error, k-fold applications of a unitary operation (but generalized to a CPTP map to account for decoherence), and measurement errors, respectively. The first line of Eq.~(\ref{eq:gk}) gives a clear experimental procedure to construct the signal functions: preparing an eigenstate of $P_\mu$ operator, apply a target unitary $k$ times, and perform projective measurement in the $P_\mu$-diagonal basis. Finally, this procedure is repeated for different eigenstate of $P_\mu$ and for different $\mu$ index as indicated in the summation appearing on the right-hand side of the first line of Eq.~(\ref{eq:gk}).  

As we focus on unital channels, the second line of Eq.~(\ref{eq:gk}) immediately follows.   $R_{\text{meas}}$ and $R_{\text{prep}}$ are the R-submatrix of Pauli transfer matrix for $\mathcal{N}_{\text {m }}$ and $\mathcal{N}_{\text {p}}$, respectively. The R-submatrix of a target unitary  $R=VDV^{-1}$ can be diagonalized. Under the trace operation, we may shuffle $V$ and $V^{-1}$ to obtain  $A_{\text{SPAM}}=V^{-1}R_{\text{meas}}R_{\text{prep}}V$, which captures all SPAM errors, in the third line.  As $D^k$ is a diagonal matrix with entries $\lambda_j^k$  we end up with a simple interpretation of $g(k)$, i.e. it is proportional to $\sum_{j=1}^N A_j \lambda_j^k$.
By using the matrix-pencil method to post-process the time series of $\{g(1),\cdots,g(K)\}$ , one may reliably extract $\lambda_j$.

\subsection{Pauli channels }
The family of Pauli channels represents a wide class of noise processes including several prominent decoherence models for quantum computations, such as the depolarizing, dephasing, bitflip and amplitude damping channels. A Pauli channel can be described by a Pauli map \cite{king2001minimal,landau1993birkhoff}. For one-qubit cases, a general Pauli channel assumes the following form,
\begin{equation}  \label{eq:pauli}
\Lambda[\rho]=\sum_{\alpha=\{0,x,y,z\}} f_{\alpha} P_{\alpha} \rho P_{\alpha}
\end{equation}
where \(P_{0}=\mathbb{I}_{2}\) and \(\left\{P_{\alpha}\vert \alpha=x,y,z \right\}\) are the Pauli operators. The coefficients \(f_\alpha \)  collectively satisfy a simple relation \(\sum_{\alpha=\{0,x,y,z\}}f_{\alpha}=1\). These maps have eigenvalues \(\lambda _\alpha\) as defined by
\begin{equation}
\Lambda\left[P_{\alpha}\right]=\lambda_{\alpha} P_{\alpha}
\end{equation}
with \(\lambda_0 = 1\). There is a simple relation between \(f_\alpha\) and \(\lambda_\alpha\),
namely,
\begin{equation}
\lambda_{\alpha}=\sum_{\beta=\{0,x,y,z\}}(-1)^{s([P_\beta,P_\alpha])} f_\alpha, \quad \alpha=x,y,z,
\end{equation}
where \(s([P_\beta,P_\alpha])=0\) if \([P_\beta,P_\alpha]=0\) and \(s([P_\beta,P_\alpha])=1\) otherwise.
A Pauli map is completely positive if and only if  its eigenvalues satisfy the Fujiwara-Algoet conditions \cite{fujiwara1999affine}
\begin{equation}
\left|1 \pm \lambda_{z}\right| \geqslant\left|\lambda_{x} \pm \lambda_{y}\right|.
\end{equation}

\subsection{Spectral quantum process tomography for non-Markovian dynamics}\label{sec:tomography}
The original SpecQPT reviewed above works naturally for a Markovian channel, but not directly applicable to non-Markovian ones. This is because a non-Markovian channel violates divisibility, 
\begin{equation} \label{eq:divisible}
\Lambda_{m+n}\neq\Lambda_m\Lambda_n,
\end{equation}
where $\Lambda_{m} \equiv \Lambda_{t_m}$. Recall that, under the standard SpecQPT framework, $g(k)$ signals are treated as a time series $\{t_0, t_1, \ldots, t_k, \ldots\}$ spaced with a uniform interval $\delta t$.  If the quantum channel under interrogation is a Markovian process, then $\Lambda_{t_{k}} = \Lambda_{\delta t}^k$, which is implicitly assumed in Eq.~(\ref{eq:gk}).  Non-Markovian channels no longer fit into the original framework that extracts spectral properties of a dynamical process from the signal function $g(k)$ by using the matrix pencil method. In this subsection, we propose a method to generalize the original framework for non-Markovian Pauli channels.

To address this deficiency, we re-define the signal functions. As the quantum channel is not divisible, we simply take every $\Lambda_{n}$ as an independent channel and introduce a subscript $t_n$ to the signal function, i.e.
\begin{equation}\label{eq:gk2}
g_{t_n}(k)=\frac{1}{2^n}\sum_{\mu=1}^{N} \operatorname{Tr}\left[P_{\mu} \mathcal{N}_{\text {m }} \circ \Lambda_{n}^{(k)} \circ \mathcal{N}_{\text {p }}\left(P_{\mu}\right)\right].
\end{equation}
The modified signal function carries two ``time" labels: operational time duration $t_n$ to denote the quantum channel $\Lambda_{n}$ under investigation, and logical time duration $k$ to imply the artificially constructed dynamical processes $\Lambda_{n}^k$, which are required for the matrix-pencil method to extract the spectrum of $\Lambda_{n}$ via $K$ time points of the signals $\{g_{t_n}(1), \cdots, g_{t_n}(K)\}$.

\begin{figure}[htp]
\centering
\includegraphics[width=0.48\textwidth]{fig1.pdf}
\caption{Up:a sequence of signal functions to extract dynamical spectral for Markovian noise. Down:group of sequence of signal functions needed to extract dynamical spectral for non-Markovian noise.}
\label{fig:signal function}
\end{figure}

If one is interested in characterizing a non-Markovian process up to time $t_M=M\delta t$, then one needs to construct $M$ independent sets of signal functions  $\{g_{t_n}(1),...g_{t_n}(K) \mid n=1,...,M\}$. For each set of signal functions, one applies the standard SpecTTM to deduce the time-evolved spectrum of $\Lambda_n$ from $K$ logical time points,
\begin{equation}
\big\{\{\lambda_1^{(1)},\lambda_2^{(1)},\ldots\}, \{\lambda_1^{(2)},\lambda_2^{(2)},\ldots\},\cdots,
\{\lambda_1^{(M)},\lambda_2^{(M)},\ldots\}\big\},
\end{equation}
where $\lambda_m^{(n)}$ denotes the $m$-th eigenvalues for $\Lambda_n$.  A number of useful properties can be inferred from these spectra, such as the degress of non-Markovianity and noise power spectrum as elucidated later in the text. See Fig.\ref{fig:signal function} for a brief summary.

\begin{figure}[htp]
\centering
\includegraphics[width=0.48\textwidth]{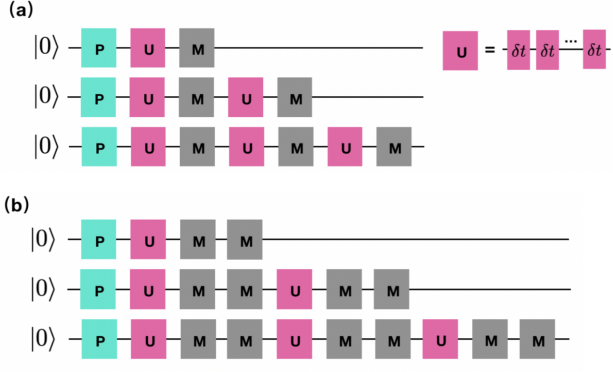}
\caption{The protocol to capture signal functions for non-Markovian noise. ``$\mathbf{P}$" stands for state preparation of eigenstate of Pauli matrix. ``$\mathbf{U}$" is the concerned quantum channel corresponding to dynamical map $\Lambda_{n}$ of various evolution time $t_n$. ``$\mathbf{M}$" is the projection measurement in the same direction of initial Pauli matrix. Up: alternately apply concerned quantum channel and projection measurement. Down: alternately apply concerned quantum channel and double projection measurements.}
\label{fig:tomography}
\end{figure}

While it is obvious that $g_{t_n}(k)$ in Eq.~(\ref{eq:gk2}) is a generalization of Eq.~(\ref{eq:gk}) for non-Markovian processes, it is not easy to experimentally construct $g_{t_n}(k)$. Next, we propose an experimental protocol to extract spectra of non-Markovian $\Lambda_{n}$ spanning a duration $T=n\delta t$. The proposed protocols requires us to perform two sets of experiments, delineated in Panels (a) and (b) of Fig.~\ref{fig:tomography}), respectively.  In this figure, symbol $\mathbf{U}$ denotes the targeted quantum channel, $\mathbf{P}$ denotes the state preparation circuit to create an eigenstate of an n-qubit Pauli matrix $P_\mu$, and $\mathbf{M}$ corresponds to a projective measurement in the eigenbasis of $P_\mu$.  

Panel (a) in Fig.~\ref{fig:tomography} implies a set of simple circuits in which $\mathbf{U}$ and $\mathbf{M}$ are interleaved for $k=1,2,3,\ldots$ times. If one repeats the experiments in Panel (a) to cover all possible initial state preparations and measurements in the Pauli basis, then one may construct a signal function, 
\begin{equation}   \label{eq:gk31}
\begin{aligned}
g^\prime_{t_n}(k)&=\frac{1}{2^n} \sum_{\mu=1}^{N} \operatorname{Tr}\left[P_{\mu}\left(\mathcal{N}_{m} \circ \Lambda_{n}\right)^{k} \circ \mathcal{N}_{p}\left(P_{\mu}\right)\right] \\
&=\operatorname{Tr}\left[(R_{\text {meas }} R_{\text{dyn}})^{k} R_{\text {prep }}\right] \\
&=\operatorname{Tr}\left[A_{\mathrm{P}} D^{k}\right]=\sum_{j=1}^{N} A_{j} (m_j\lambda_{j}^{(n)})^{k},
\end{aligned}
\end{equation}
where $R_{\text{meas}}$, $R_{\text{dyn}}$, and $R_{\text{prep}}$ are the sub-matrix $R$ of the Pauli transfer matrix for $\mathcal{N}_{\text {meas }}$, $\Lambda_{n}$ and $\mathcal{N}_{\text {prep}}$, respectively.  $R_{\text{meas}}R_{\text{dyn}}=VDV^{-1}$ can be diagonalized with eigenvalues $\tilde{\lambda}_j^{(n)}=m_{s(j)} \lambda_{j}^{(n)}$, where $m_s$ and $\lambda_j^{(n)}$ are eigenvalues of $R_{\text{meas}}$ and $R_{\text{dyn}}$, respectively. This special relation (spectra of a matrix product is derived from multiplications of the spectrum of two matrices)
holds for Pauli channels. $A_{P}=V^{-1}R_{\text{prep}}V$ encapsulates all the state-preparation errors. As every application of dynamical map $U$ is followed with a projective measurement, the fundamental components that scale with $k$ are actually $(m_{s(j)}\lambda_{j}^{(n)})^{k}$.  Hence, applying the matrix-pencil method to the time series constituted by K signal functions, $g^\prime_{t_n}(k)$, would give us an estimate of $m_{s(j)}\lambda_{j}^{(n)}$ instead.

To isolate $\lambda_{j}^{(n)}$, we need additional data.  Hence, we propose to perform a second set of experiments delineated in panel (b) of Fig.\ref{fig:tomography}.  In this case, the experiment protocol is almost identical to the previous experiment except that every application of a dynamical map $U$ is followed by two consecutive projective measurements.  Now, one may construct a new set of signal functions,
\begin{equation}   \label{eq:gk41}
\begin{aligned}
g^{\prime\prime}_{t_n}(k)&=\frac{1}{2^n} \sum_{\mu=1}^{N} \operatorname{Tr}\left[P_{\mu}\left(\mathcal{N}_{m} \circ \mathcal{N}_{m} \circ \Lambda_{n}\right)^{k} \circ \mathcal{N}_{p}\left(P_{\mu}\right)\right]   \\
&=\operatorname{Tr}\left[(R_{\text {meas }} R_{\text {meas }} R_{\text{dyn}})^{k} R_{\text {prep }}\right]  \\
&=\operatorname{Tr}\left[A_{\mathrm{P}} D^{k}\right]=\sum_{j=1}^{N} A_{j} (m_j^2\lambda_{j}^{(n)})^{k}.
\end{aligned}
\end{equation}
It is clear that the matrix pencil method may give us an estimate of $(m^2_{s(j)}\lambda_{j}^{(n)})$, which are eigenvalues of $R^2_{\text {meas }} R_{\text{dyn}}$.  Now, by combining the experimental results of these two sets of experiments, we can easily obtain
\begin{equation}\label{eq:lambda}
\lambda_{j}^{(n)}=\frac{\left(m_{j} \lambda_{j}^{(n)}\right)^{2}}{m_{j}^{2} \lambda_{j}^{(n)}}
\end{equation}
If we perform these two sets of experiments, panels (a) and (b) of Fig.~\ref{fig:tomography}, to deduce the time-evolved spectra, $\lambda_{j}^{(n)}$ for $n=1,\cdots, M$ without SPAM errors.

\subsection{Spectral transfer tensor method }\label{sec:specTTM}
We now introduce spectral transfer tensor map (SpecTTM) for Pauli channels. We will discuss how the method can be useful for more general dynamical processes in Sec.\ref{sec:optimal-basis}. 

Since Pauli channels necessarily have a diagonal Pauli transfer matrix in a fixed basis for all time, it can be proved inductively that $T_n$ derived from Eq.~(\ref{eq:ttm}) must be diagonal in the same basis. The non-Markovian SpecQPT described in the previous subsection can be invoked to give diagonal elements for $\Lambda_{m}$. Via Eq.~(\ref{eq:ttm2}), one may recursively derive the following relations,
\begin{subequations}
\begin{align}
\lambda_{n}^{\alpha}=\sum_{m=0}^{n-1} \tau_{n-m}^{\alpha} \lambda_{m}^{\alpha} \label{eq:specttma},  \\
\tau_{n}^{\alpha}=\lambda_{n}^{\alpha}-\sum_{m=1}^{n-1} \tau_{n-m}^{\alpha} \lambda_{m}^{\alpha},   \label{eq:specttmb}
\end{align}
\end{subequations}
where $\tau_{n}^{\alpha}$ denotes an eigenvalue of $T_n$. In this work, we refer to the set $\{ \tau_n^{\alpha} \mid \alpha \in \{x,y,z\} \text{ and } n=1,\ldots,M\}$ as the SpecTTM.  As discussed before, it is often appropriate to assume the memory kernel of a non-Markovian process has finite width and truncate the summation in Eq.~(\ref{eq:specttma}) by keeping the first $M$ terms and discardnig the rest.

\subsection{Resource consumption}
Finally, we analyze the efforts required to implement SpecTTM based on the protocols introduced in Fig.~\ref{fig:tomography}.  According to Eqs.~(\ref{eq:gk31})-(\ref{eq:gk41}), each of these two modified signal functions requires the same amount of experimental efforts as the original SpecQPT. Since SpecTTM requires a characterization of $M$ dynamical maps to cover the memory kernel, there is a total of $d \times (d^2-1) \times (K+1) \times M$ distinct experiments (i.e. quantum circuits). Each distinct experimental setup should be further repeated  $N_{\text{samples}}$ times in order to reach an acceptable variance $\propto \mathcal{O}(1/N_{\text{samples}})$ for the measurement statistics. While the present approach is not scalable, it is actually a very economical approach to obtain the SPAM-free estimate of a memory kernel for a non-Markovian Pauli channel, if we compare to a direct application of SPAM-free process tomographic technique, such as the gate-set tomography, to derive TTM for the same quantum system, which require about $d^2\times(d^2-1)\times M$ distinct experiments.


\section{Noise Spectroscopy based on SpecTTM }\label{sec:noise}
In this section we further discuss how SpecTTM can be used to (1) quantify non-Markovianity and (2) reconstruct power spectrum for Pauli channels. Before delving into these tasks, we first relate the spectral properties of a Pauli channel to the corresponding time-local and non-local master equations.

\subsection{Master equations for Pauli channels}\label{sec:paulimaster}
%

Dynamics for every open quantum system can be described by the Nakajima-Zwanzig equation,
\begin{equation}\label{eq:nzeq}
\dot{\rho}_{t}=\int_{0}^{t} \mathcal{K}_{t-t'}, \rho_{t'}, d t^{\prime}
\end{equation}
where $\mathcal{K}_t$ is the non-Markovian memory kernel that relates a system's past dynamical history to its present evolution under the influence of an external environment such as noise. In addition to Eq.~(\ref{eq:nzeq}), the dynamical evolution can be alternatively represented in a time-local form as follows,
\begin{equation} \label{eq:L}
\dot{\rho}_t = \left(\int_0^t \mathcal{K}_{t-t})\Lambda_{t'}\Lambda^{-1}_tdt'\right)\rho(t)\equiv\mathcal{L}_t\rho(t).
\end{equation}

For Pauli channels $\Lambda_t$, that is diagonal in the Pauli basis, one immediately infers that both $\mathcal{L}_t$ and $\mathcal{K}_t$ are also diagonal in the Pauli basis for all time $t$.  In fact, one can explicitly show that
\begin{eqnarray}
\mathcal{K}_t  & = & \sum_{\alpha={x,y,z}} k_\alpha(t)[\mathbb{U}_\alpha-\mathbb{I}], \nonumber \\
\mathcal{L}_{\mathrm{t}} & =&\frac{1}{2}\sum_{\alpha={x,y,z}} \gamma_{\alpha}(t) \left[\mathbb{U}_{\alpha}-\mathbb{I}\right],
\end{eqnarray}
where $\mathbb{U}_{\alpha}[\rho]=P_{\alpha} \rho P_{\alpha}^{\dagger}$ and $\mathbb{I}[\rho] = \rho$.   Hence, the dynamical evolution of a Pauli channel is also fully encapsulated in the parameters $k_\alpha(t)$ and $\gamma_{\alpha}(t)$.  To quantify non-Markovianity and reconstruct the noise power spectrum of a non-Markovian dynamics, it is beneficial to have $k_\alpha(t)$ and $\gamma_{\alpha}(t)$ readily extracted from experiments. 

Casting the Nakajima-Zwanzig equation $\dot{\rho}\left(t_{n}\right)=\sum_{m=0}^{n-1} \mathcal{K}_{n-m} \rho\left(t_{m}\right) \delta t$ in a discretized form, one may relate the the memory kernel to the TTMs in Eq.~(\ref{eq:ttm2}) if these maps are sampled at sufficiently small time step $\delta t$,
\begin{equation}  \label{eq:KT1}
T_{n}=\mathcal{K}_{n} \delta t^{2}+\delta_{n, 1}.
\end{equation}
Noting the TTM and the memory kernel in Eq.~(\ref{eq:KT1}) should assume a diagonal form for Pauli channels, we immediately obtain 
\begin{equation}
k^{\alpha}_n \delta t^2 = \tau_n^\alpha  - \delta_{n,1}.
\end{equation}
This relation above indicates that one may readily reconstruct the memory kernel from SpecTTM with minimal efforts.

Next, the generator for a Pauli channel must satisfy
\begin{equation}
\mathcal{L}_{t}\left[P_{\alpha}\right]=\mu_{\alpha}(t) P_{\alpha},
\end{equation}
with $\mu_{\alpha}(t)=\gamma_{\alpha}(t)-\sum_{\beta=\{x,y,z\}} \gamma_{\beta}(t)$. From Eq.~(\ref{eq:L}), we may deduce
\begin{eqnarray}
\dot{\Lambda}_{\mathrm{t}}\left[P_{\alpha}\right]=\mathcal{L}_{t} \Lambda_{t}\left[P_{\alpha}\right],
\end{eqnarray}
and draw the conclusion that $\lambda_{\alpha}(t)=\operatorname{Exp}\left[\int_{0}^{t} \mu_{\alpha}(\tau) d \tau\right]$. Thus, we acquire a relationship between $\gamma_\alpha(t)$ and $\lambda_\alpha$ as follows,
\begin{equation} \label{eq:decoherencerate}
\Gamma_t = \int_{0}^{t} \gamma_{\alpha}(\tau) d \tau=\frac{1}{2} \ln \frac{\lambda_{\alpha}(t)}{\lambda_{\beta}(t) \lambda_{\eta}(t)}, \quad \alpha \neq \beta \neq \eta.
\end{equation}
With help of SpecTTM via Eq.~(\ref{eq:specttma}), we may estimate $\Gamma_t$ for all time $t$.

\subsection{RHP measure for Non-Markovianity}\label{sec:RHP}
A mathematically rigorous measure for non-Markovianity of a dynamical process was proposed by Rivas, Huelga, and Plenio \cite{rivas2010entanglement}.  This measure is motivated by the fact that non-Markovian dynamical maps are not CP-divisible.  Since a local CP-divisible map must maintain the positivity of a density matrix $\rho=|\Phi\rangle\langle\Phi|$ (where $|\Phi\rangle$ is a maximally entangled state of a quantum system and an ancillary counterpart), the violations of the positivitiy of $\Lambda_{t}\otimes I[\rho]$ gives us one way to quantify non-Markovianity. More precisely, the RHP measure is given by
\begin{equation}
\mathcal{I}=\int_{0}^{\infty} \mathbf{g}(t) d t
\end{equation}
with
\begin{equation}
\mathbf{g}(t)=\lim _{\epsilon \rightarrow 0^{+}} \frac{\|\left[I+\left(\mathcal{L}_{t} \otimes I\right) \epsilon\right]|\Phi\rangle\langle\Phi| \|_1-1}{\epsilon}
\end{equation}
where  $\vert\vert \cdot \vert\vert_1$ denotes the trace norm, $\mathcal{L}_{t}$ is the generator of $\Lambda_t$ as defined in Eq.~(\ref{eq:L}).  For Pauli channels, a straightforward calculation leads to
\begin{equation}
\mathbf{g}_\alpha(t)=\left\{\begin{array}{ccc}0 & \text { for } & \gamma_\alpha(t) \geqslant 0 \\ -\gamma_\alpha(t) & \text { for } & \gamma_\alpha(t)<0\end{array}\right.
\end{equation}
and finally
\begin{equation}
\mathcal{I}=\sum_{\alpha=\{x,y,z\}}{\mathcal{I_\alpha}}=\sum_{\alpha=\{x,y,z\}}{\int_{\gamma_\alpha(t)<0} -\gamma_\alpha(t) d t}
\end{equation}
Thus, an accumulation of negative decoherence rates $\gamma_\alpha(t)$ over the duration $t \in [0,\infty]$ accounts for non-Markovianity. According to Eq.~(\ref{eq:decoherencerate}), decoherence rates are related to the spectra of dynamical maps  that can be deduced from SpecTTM for arbitrary long time.

\subsection{Noise Power Sepctrum}\label{sec:correlation}
Next we describe how the SpecTTM allows one to reconstruct the noise power spectrum for an open quantum system.  Let us consider a set of qubits governed by the following Hamiltonian,
\begin{eqnarray}\label{eq:H1}
H(t) & = & H_s + H_{sb}(t) \nonumber \\
& = & H_s + \sum_{i,\alpha} g_{i} B^\alpha_i(t) \sigma^\alpha_i,
\end{eqnarray}
where $H_s$ is a time-independent system Hamiltonian for the qubits, and $H_{sb}$ denotes the system-noise interaction. $\sigma_i^\alpha$ is a Pauli operator with the index $i$ labeling the qubits and the index $\alpha$ one of the $\{x,y,z\}$ Cartesian components. $B^\alpha_i(t) = e^{-i H_b t}B^\alpha_i(0)e^{i H_b t}$ is a bath operator in the interaction picture with respect to $H_b$, the environment Hamiltonian.  If assuming Gaussian noises, then the environment-induced perturbations can be fully captured in the first two statistical moments $\langle B^\alpha_i(t) \rangle$ and $\langle B^\alpha_i(t) B^\beta_j(t') \rangle$ with $\langle \cdot \rangle$ implying an average with $\rho_b$, such as the thermal state for the bath. 

For a piece of high-quality quantum hardware, it is often sufficient to consider the weak-coupling regime for system-noise coupling. The exact memory kernel \cite{breuer_theory_2007} for an arbitrary quantum bath can be written as follows,
\begin{eqnarray}\label{eq:memker}
\mathcal{K}(t, t^\prime) & = & \mathcal{P}\mathcal{L}(t)\exp_+\left[\int^t_{t^\prime} ds \mathcal{QL}(s)\right]\mathcal{QL}(t^\prime) \mathcal{P},
\end{eqnarray}
where projection operators $\mathcal{P}$ and $\mathcal{Q}=1-\mathcal{P}$ are defined by $\mathcal{P}\Omega(t)=\text{Tr}_b(\Omega(t))\otimes\rho_b$, i.e. $\mathcal{P}$ projects a system-bath entangled quantum state $\Omega(t)$
to a factorized form consisting of a system part $\rho(t)=\text{Tr}_b \Omega(t)$, and a bath part $\rho_b$ which is a stationary state with respect to the bath Hamiltonian $H_b$. We note that the kernel is a stationary process $\mathcal{K}(t,t^\prime)=\mathcal{K}(t-t^\prime)$ when the noise satisfies the stationary Gaussian conditions. 

The memory kernel for a Gaussian noise can be expressed as $\mathcal{K}(t)=\sum_{n=1}^\infty \mathcal{K}_{2n}(t)$, where $\mathcal{K}_{2n}(t)$ corresponds to the order $2n$ expansion of the Hamiltonian with respect to $H_{sb}$. Keeping only the leading (i.e. second order) term  gives
\begin{eqnarray}\label{eq:corr}
 \mathcal{K}(t)(\boldsymbol{\cdot}) & \approx & \mathcal{K}_2(t)  \\
 & = &  \sum_{\alpha\alpha^\prime} [\sigma^\alpha, C_{\alpha\alpha^\prime}(t) \sigma^{\alpha^\prime}(t) (\boldsymbol{\cdot} )
- C^*_{\alpha\alpha^\prime}(t)(\boldsymbol{\cdot}) \sigma^{\alpha^\prime}(t) ], \nonumber 
\end{eqnarray}
where $\sigma^{\alpha}(t)=\exp(-iH_{s}t)\sigma^\alpha \exp(iH_s t)$ and the bath correlation functions are given by
\begin{eqnarray}\label{eq:corrdef}
C_{\alpha\alpha^\prime}(t)= g^2 \left\langle \hat{B}^\alpha(t) \hat{B}^{\alpha^\prime}(0) \right\rangle.
\end{eqnarray}
We suppress the index $i$ on Pauli matrices and noise operators $\hat B^\alpha(t)$in Eqs.~(\ref{eq:corr})-(\ref{eq:corrdef}), as we should illustrate the idea with an one-qubit case from here onward. 

For the Pauli channel defined in Eq.~(\ref{eq:pauli}), the weak-coupling (second-order approximated) memory kernel reads
\begin{equation} \label{eq:2th}
\mathcal{K}(t) \rho(t)=\sum_{\alpha={x,y,x}}\left[C_{\alpha \alpha}(t)+C_{\alpha \alpha}^{*}(t)\right] \left(\sigma_{\alpha} \rho(t) \sigma_{\alpha}-\rho(t)\right).
\end{equation}
On the other hand, the memory kernel of a Pauli channel is known to possess a simple expression,
\begin{equation} \label{eq:Kernel}
\mathcal{K}_{t}=\sum_{\alpha={x,y,z}} \frac{1}{2}
k_{\alpha}(t)\left[\mathbb{U}_{\alpha}-\mathbb{I}\right].
\end{equation}
Relating Eq.~(\ref{eq:2th}) and (\ref{eq:Kernel}), we identify a way to reconstruct $C_{\alpha\alpha}(t)$ from SpecTTM data (via first calculating $k_\alpha(t)$ as discussed in a previous section),
\begin{equation} \label{eq:correlation}
\begin{aligned}
k_{\alpha}\left(t_{n}\right)=&2\left[C_{\alpha \alpha}\left(t_{n}\right)+C_{\alpha \alpha}^{*}\left(t_{n}\right)\right]  \\
&-\sum_{\beta={x,y,x}} 2\left[C_{\beta \beta}\left(t_{n}\right)+C_{\beta \beta}^{*}\left(t_{n}\right)\right]
\end{aligned}
\end{equation}
Once the noise correlation function is determined, the corresponding spectral density can be determined by invoking the fluctuation-dissipation theorem \cite{yan_xu_annphyschem05}, which gives
\begin{eqnarray} \label{eq:omega}
J_{\alpha\alpha}(\omega) &=& \frac{1}{2} \int^\infty_{-\infty} dt e^{i\omega t}\left[ C_{\alpha\alpha}(t)-C^*_{\alpha\alpha}(t) \right].
\end{eqnarray}

\section{Beyond Pauli Channel}\label{sec:nopauli}
In the previous sections, we introduce SpecTTM for non-Markovian Pauli channels. For more general cases, we apply the Pauli twirling approximation (PTA) in order to extract useful information on the non-Markovian noisy process. 

\subsection{Pauli twirling approximation }\label{sec:PTA}

According to Ref.\cite{geller2013efficient}, a quantum channel $\Lambda$  can be approximated by a Pauli channel $\tilde{\Lambda}$ through Pauli twirling approximating (PTA),
\begin{eqnarray}
\tilde{\Lambda}(\rho) &=&\frac{1}{K} \sum_{v=0}^{K-1} P_{v} \Lambda\left[P_{v} \rho P_{v}\right] P_{v} \nonumber \\
& = &\sum_{P_{v} \in \mathcal{P}_{n}} f_{vv} P_{v} \rho P_{v}.
\end{eqnarray}
We apply PTA to the spectral quantum process tomography elaborated in Sec.\ref{sec:tomography} to extract SPAM free  spectral of dynamical map for twirled Pauli channel.  Under the PTA, the generalized signal functions for non-Markovian processes, defined in Eq.~(\ref{eq:gk31}) and Eq.~(\ref{eq:gk41}), should read
\begin{equation}\label{eq:gkpta}
\begin{aligned} g^\prime_{t_{n}}(k)=&\frac{1}{2^n}  \sum_{\mu=1}^{N} \operatorname{Tr}\left[P_{\mu} \mathcal{N}_{m} \circ \mathcal{N}_{t} \circ \Lambda_{n} \circ \mathcal{N}_{t} \right.\\ & \circ\left(M_{\mu} \circ \mathcal{N}_{m} \circ \mathcal{N}_{t} \circ \Lambda_{n} \circ \mathcal{N}_{t}\right)^{k-1} \\ &  \circ \mathcal{N}_{p}\left(P_{\mu}\right)] \end{aligned}, 
\end{equation}
and
\begin{equation}\label{eq:gkpta2}
\begin{aligned} g_{t_{n}}^{\prime\prime}(k)=& \frac{1}{2^n} \sum_{\mu=1}^{N} \operatorname{Tr}\left[P_{\mu} \mathcal{N}_{m} \circ M_{\mu}^{\prime} \circ \mathcal{N}_{t} \circ \Lambda_{n} \circ \mathcal{N}_{t}\right.\\ & \circ\left(M_{\mu}^{\prime} \circ M_{\mu}^{\prime} \circ \mathcal{N}_{t} \circ \Lambda_{n} \circ \mathcal{N}_{t}\right)^{k-1} \\ & \circ \mathcal{N}_{p}\left(P_{\mu}\right)] \end{aligned},
\end{equation}
where $\tilde{\Lambda}=\mathcal{N}_{t} \circ \Lambda \circ \mathcal{N}_{t}$ is the twirled Pauli channel. As implicitly indicated in the equations above, one should apply a twirling approximation between every operations in the non-Markovian SpecQPT protocol in \textcolor{red}{section} . 

As proved in Appendix \ref{app:signalfunction}, the projective measurements based spectral tomographic technique introduced in Sec.\ref{sec:tomography} is compatible with PTA. Hence, from $g^\prime_{t_n}(k)$ and $g^{\prime\prime}_{t_n}(k)$, one obtains the spectrum for the twirled Pauli channel $\tilde{\Lambda}$.



\subsection{Optimal twirling basis for noise characterization}\label{sec:optimal-basis}
In the previous section, the aforementioned Pauli twirling approximation is implicitly assumed to be done in the computational basis. Hence, one obtains a partial (and potentially skewed) spectral information of the original quantum channel $\Lambda$.  While one may speculate actual properties of a noisy quantum process from results obtained under a PTA, we show that the degree of non-Markovianity (based on the RHP measure) may be more reliably estimated in an optimal basis defined below. 

For  simplicity,  we  illustrate the idea of PTA in an optimal basis with a single qubit. First, we introduce a new basis $P_v^{\prime}$ by applying some unitary transformation to the standard Pauli basis,
\begin{equation}
P_{v}^{\prime} \in\left\{I, U P_{x} U^{\dagger}, U P_{y} U^{\dagger}, U P_{z} U^{\dagger}\right\}
\end{equation}
where $U\left(\theta_{1}, \theta_{2},\theta_{3}\right)=\mathcal{R}_{\hat{z}}\left(\theta_{1}\right) \mathcal{R}_{\hat{y}}\left(\theta_{2}\right)\mathcal{R}_{\hat{z}}\left(\theta_{3}\right)$ and $\mathcal{R}_{\hat{n}}(\theta)$ denotes rotating an angle of $\theta$ around the $\widehat{n}$  axis. A PTA channel in the new basis assumes the following form,
\begin{equation}
\tilde{\Lambda}(\rho)=\sum_{P'_{v} \in \mathcal{P}_{n}} f'_{v v} P'_{v} \rho P'_{v}.
\end{equation}

Once, a new Pauli basis has been determined, we proprose to generate SpecTTM data by first preparing the modified signal functions introduced in Eqs.~(\ref{eq:gkpta})-(\ref{eq:gkpta2}) with the understanding that PTA is done in this new basis (given by $U\left(\theta_{1}, \theta_{2},\theta_{3}\right)$) that maximizes the RHP measure for the PTA map $\tilde{\Lambda}_\theta$. 

\section{Results }\label{sec:simulation}
We illustrate how to use SpecTTM to quantify non-Markovianity and noise power spectrum for Pauli and non-Pauli channels through numerical experiments. 

\subsection{SpecTTM spectroscopy of Pauli channel }\label{sec:sim-pauli}
We first consider an one-qubit pure phasing model, which is a Pauli channel. The Hamiltonian reads $H_s=\omega_{s}\sigma^z$ and $H_{sb}=B^z(t)\sigma^z$. The noise $B^z(t)$, a Gaussian process, possesses the statistical moments: $\langle B^z(t) \rangle = 0$ and $C_{zz}(t-t^\prime)=\langle B^z(t)B^z(t^\prime)\rangle = \lambda e^{-\left|t-t'\right|}\cos[\omega_{c}(t-t')]$. For these cosine functions modulated by an exponentially decaying envelope, the corresponding power spectrum assumes a Lorentzian profile. 

\begin{figure}[htp]
\centering
\includegraphics[width=0.46\textwidth]{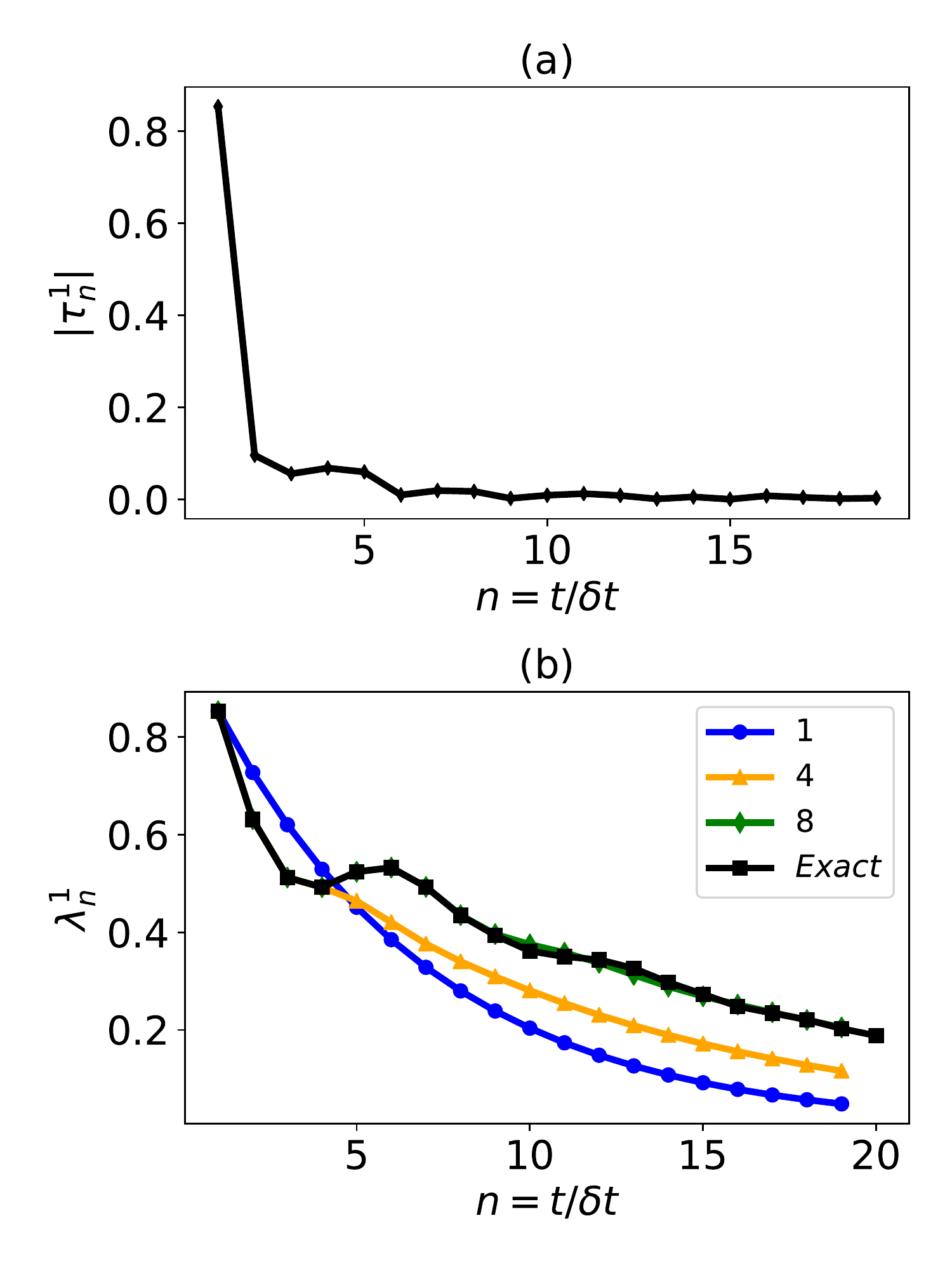}
\caption{SpecTTM for a single-qubit pure dephaing model. (Top): absolute value distribution of the SpecTTM, $\tau_n^\alpha$ over time.  (Bottom): dynamical map spectral predictions by SpecTTM. The model parameters: $H_{sb}=B^z(t)\sigma ^z$, $C_{zz}(0)=\lambda=4$ and $\delta t=0.2$.}
\label{fig:toy-ttmnorm-predict}
\end{figure}

The pure-dephasing dynamical maps are given by
\begin{eqnarray}\label{eq:toy-dmap}
\Lambda_n = \left[\begin{array}{cccc}
1 & 0 & 0 & 0 \\
0 & e^{-\Upsilon(t_n)+i2\omega_s t} & 0 & 0 \\
0 & 0 & e^{-\Upsilon(t_n)-i2\omega_s t} & 0 \\
0 & 0 & 0 & 1 \end{array} \right],
\end{eqnarray}
with
\begin{eqnarray}\label{eq:gammat}
\Upsilon(t)=\frac{4}{\pi}\int^\infty_0 d\omega\frac{S(\omega)}{\omega^2}[1-\cos(\omega t)].
\end{eqnarray}
Note that these dynamical maps satisfy  $\rho(t_n) = \Lambda_n \rho(0)$ as required.
It is straightforward to verify that these maps are not divisible, i.e. $\Lambda_{n+m}\neq \Lambda_n\Lambda_m$, and the dynamics is clearly non-Markovian.  The spectrum of the dynamical map may be read off the diagonal in Eq.~(\ref{eq:toy-dmap}) as $\{ e^{-\Upsilon(t_n)+i2\omega_s t},e^{-\Upsilon(t_n)-i2\omega_s t},1\}$.

\begin{figure}[hbt]
\centering
\includegraphics[width=0.48\textwidth]{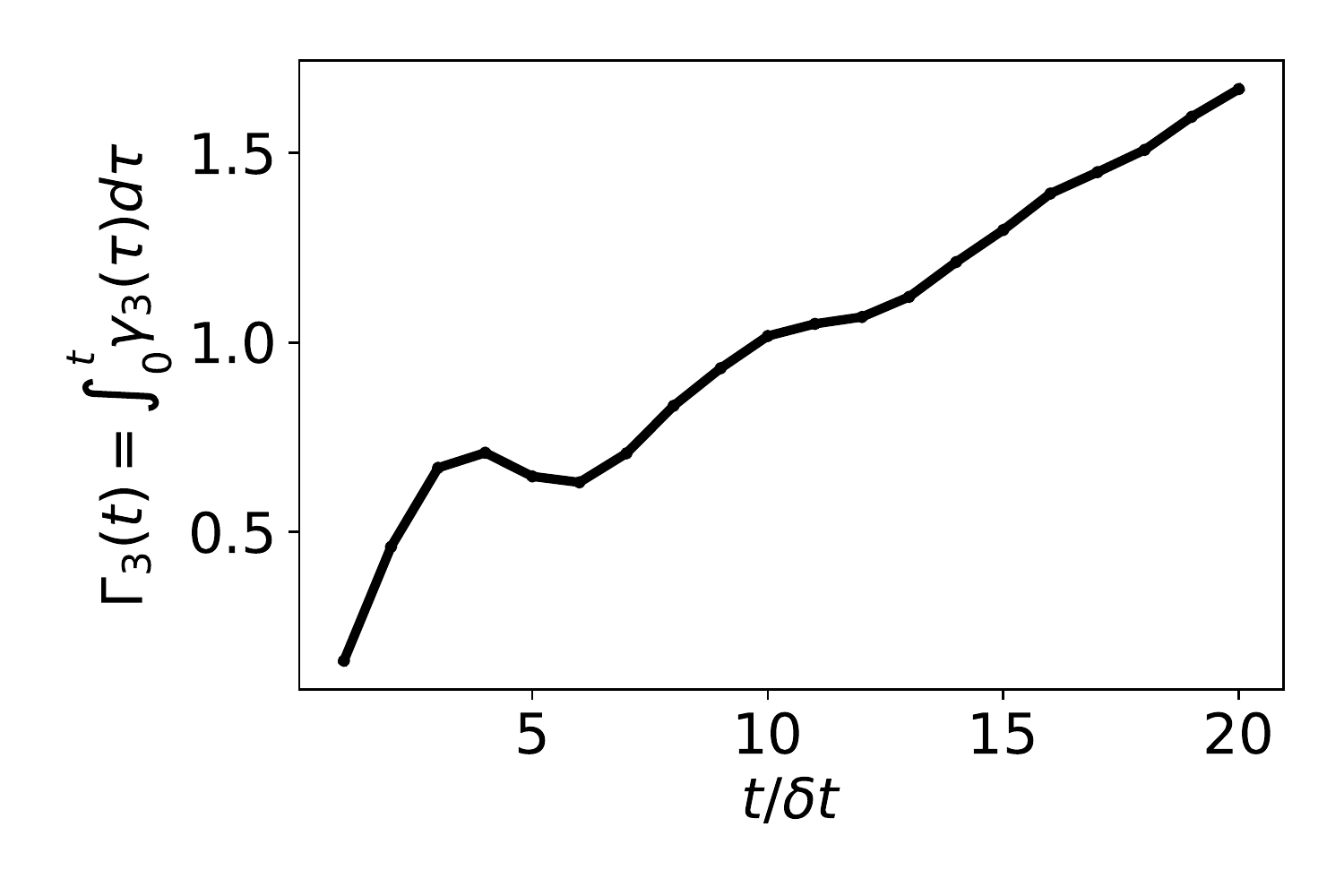}
\caption{Non-Markovianity of single-qubit pure dephasing model. The model parameters: $H_s=0.1\sigma_z$, $H_{sb}=B^z(t)\sigma ^z$, $C_{zz}(0)=\lambda=4$ and $\delta t=0.2$.}
\label{fig:toy-measure}
\end{figure}

We complicate the simulation of this one-qubit pure dephasing model by adding random noises during the state preparations and measurements. We then extract the SPAM error-free eigenvalues of the dynamical maps with the SpecQPT and construct the SpecTTM according to  Eq.~(\ref{eq:specttma}). In the upper panel of Fig.~\ref{fig:toy-ttmnorm-predict} we plot $\vert \tau^\alpha_n \vert$ at different times. Since, for the pure dephasing model, $\vert \tau^1_n \vert=\vert \tau^2_n \vert$, $\vert \tau^3_1 \vert=1$ and $\vert \tau^3_n \vert=0$ for $n>1$, we only present the result for $\vert \tau^1_n \vert$. As illustrated in the upper panel of Fig.~\ref{fig:toy-ttmnorm-predict}, more than one TTM have non-neligibile norm. This observation confirms that the non-Markovianity of a dynamical process is directly correlated with $\vert \tau^\alpha_n \vert$ distribution. Furthermore, it is clear that the higher order $\vert \tau^\alpha_n \vert$ are increasingly suppressed. This trend justifies an earlier claim that we should be able to truncate Eq.~(\ref{eq:specttmb}) to only the first few SpecTTMs with non-trivial norms for dynamical predictions based on Eq.~(\ref{eq:specttma}). In the lower panel of Fig.~\ref{fig:toy-ttmnorm-predict}, we plot the spectrum of the dynamical map as a function of time.  Since $\vert \lambda^1_n \vert=\vert \lambda^2_n \vert$ and $\vert \lambda^3_n \vert=1$, we only present the result for $\vert \lambda^1_n \vert$. The curve with black squares is the exact result $e^{-\Upsilon(t)}$, which requires explicitly evaluating the integral in Eq.~(\ref{eq:gammat}) at every time point.  The other curves are results obtained by using different numbers of SpectTTMs in the way prescribed by Eq.~(\ref{eq:specttma}) to predict quantum dynamical evolution. In this case, we consider using the first $n=1,4,8$ 
SpecTTMs, respectively.  Accurate results are obtained for the entire simulation duration when a sufficient number ($n=8$) of SpecTTMs are taken into account.

Next we analyze the non-Markovianity of the dynamical process by plotting $\Gamma_\alpha(t)$, the integral of decoherence rate at different times (as defined in Eq.~(\ref{eq:decoherencerate})) in Fig.~(\ref{fig:toy-measure}). As discussed in the previous section, the RHP measure for non-Markovianity implies that the non-monotonic trending behaviour of $\Gamma_\alpha(t)$ is an absolute signature of non-Markovian dynamics.  This non-monotonicity is indeed manifested around $t_4$ in Fig.~(\ref{fig:toy-measure}).  We only show $\Gamma_3(t)$, since $\Gamma_1(t)=\Gamma_2(t)=0$. This is an encouraging indication that SpecTTM may facilitate the implementation of the RHP measure in an experiment.

\begin{figure}[htp]
\centering
\includegraphics[width=0.49\textwidth]{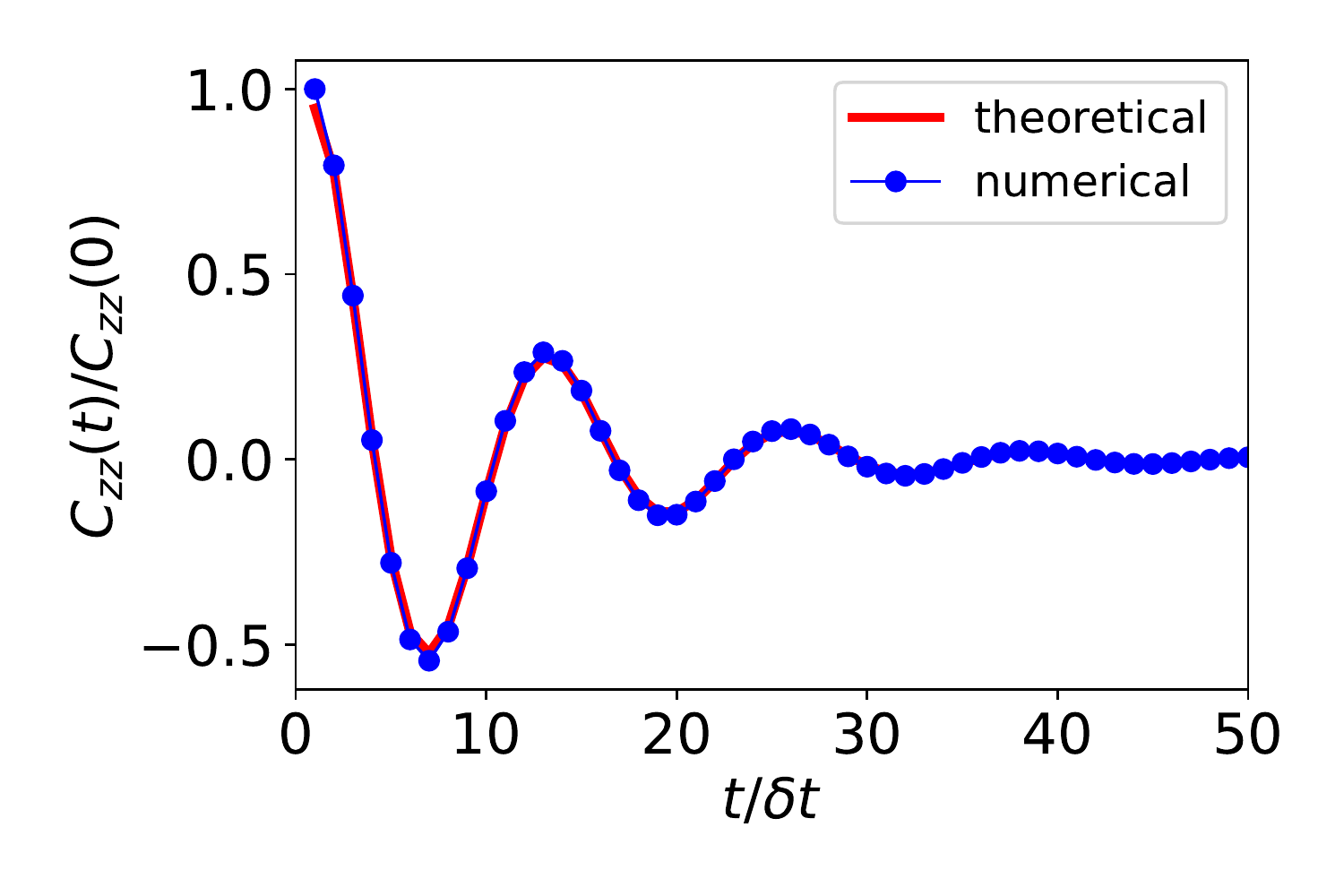}
\caption{TTM noise spectroscopy of pure dephasing model. The numerically extracted correlation function (from the memory kernel) matches well with the theoretical result $C_{zz}(t)=\langle B^z(t)B^z(0) \rangle$. Model parameters: $H_s=0.02\sigma_z$,  $C_{xx}(0)=0.04$ and $dt=0.1$.}
\label{fig:toy-specden}
\end{figure}

A critical role of a quantum noise spectroscopy is to determine the noise power spectrum.  Since a spectrum is essentially the Fourier transform of the corresponding correlation function, we will be content if we may easily obtain noise correlation functions in experiments. We consider the case of weakly coupled noise where the approximation $\mathcal{K}(t)\approx \mathcal{K}_2(t)$ is valid. We use Eq.~(\ref{eq:correlation}) to infer the targeted correlation function. In Fig.~(\ref{fig:toy-specden}), we plot the numerically recovered correlation function based on the TTM data, and it agrees well with the theoretical correlation function that the random noise $B^z(t)$ must satisfy in our numerical experiment. 

Besides the pure dephasing channel, other well known noisy dynamics such as depolarization channel and amplitude damping channel in a quantum circuit can all be rigorously analyzed with the SpecTTM (which fully characterizes non-Markovian dynamics of any Pauli channels).   While depolarization process is clearly a Pauli channel,the single-qubit amplitude damping channel deserves some clarifications. We note the amplitude damping channel can be described by he Kraus matrix,
\begin{equation}
E_1=\left(\begin{array}{cc}1 & 0 \\ 0 & \sqrt{1-p} \end{array}\right), E_2=\left(\begin{array}{cc}0 & \sqrt{p} \\ 0 & 0 \end{array}\right),
\end{equation}
It can be expressed by Pauli matrices
\begin{equation} \label{eq:ampdamp}
\Lambda(\rho)=\sum_{P_{v} \in \mathcal{P}_{n}} f_{v v'} P_{v} \rho P_{v'}
\end{equation}
with $f_{00}=(\frac{1+\sqrt{1-p}}{2})^2, f_{11}=f_{22}=\frac{p}{4},f_{33}=(\frac{1-\sqrt{1-p}}{2})^2,f_{03}=f_{30}=\frac{p}{4},f_{21}=-f_{12}=\frac{-p}{4i}$. It is clear that Eq.~(\ref{eq:ampdamp}) is not a Pauli channel. However, according to Eq. (\ref{eq:PTA}), the resulting Pauli transfer matrix is diagonal with $R_{11}=\sqrt{1-p},R_{22}=\sqrt{1-p},R_{33}=1-p$, which is the same as the results after a PTA. This is why SpecTTM can accurately characterize non-Markovianity of an amplitude damping channel.

\subsection{SpecTTM spectroscopy beyond Pauli channel}\label{sec:sim-nonpauli}
\begin{figure}[htp]
\centering
\includegraphics[width=0.48\textwidth]{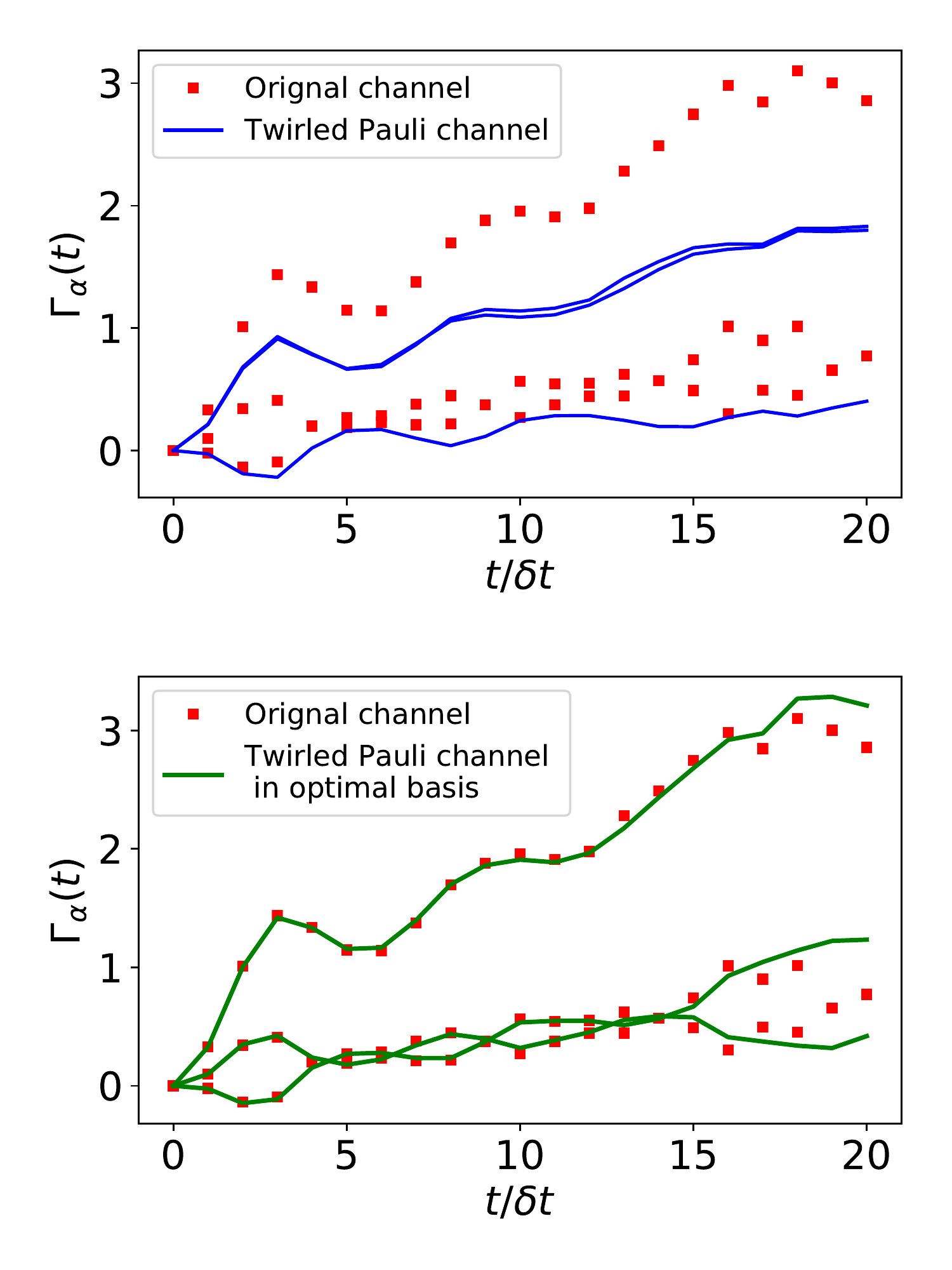}
\caption{Non-Markovanity of non-Pauli channel. (Top): the non-Markovanity of original non-Pauli channel (red squares) and non-Markovanity of Pauli channel by PTA in standard basis (blue lines) . (Bottom):  the non-Markovanity of original non-Pauli channel (red squares) and non-Markovanity of Pauli channel by PTA in optimal basis (blue lines). The model parameters:  $H_{sb}=B^x(t)\sigma ^x+B^y(t)\sigma ^y$, $C_{xx}(0)=\lambda_1=5,C_{yy}(0)=\lambda_2=5,C_{xy}(0)=\lambda_3=3$ and $\delta t=0.2$.}
\label{fig: nonPauli-RHP}
\end{figure}

We next illustrate how well SpecTTM may charcterize a non-Pauli channel. For simplicity, let us consider a general Hamiltonian for an idling qubit in a quantum circuit,
\begin{equation}
H_{sb}=B^x(t)\sigma^x+B^y(t)\sigma^y,
\end{equation}
where $B^x(t),B^y(t)$ are assumed to be correlated Gaussian noises, and satisfy the statistical moments: $\langle B^x(t) \rangle = 0, \langle B^y(t) \rangle = 0$ with $C_{xx}(t-t^\prime)=\langle B^x(t)B^x(t^\prime)\rangle = \lambda_1 e^{-\left|t-t'\right|}\cos[\omega_{c}(t-t')]$, $C_{yy}(t-t^\prime)=\langle B^y(t)B^y(t^\prime)\rangle = \lambda_2 e^{-\left|t-t'\right|}\cos[\omega_{c}(t-t')]$, and $C_{xy}(t-t^\prime)=\langle B^x(t)B^y(t^\prime)\rangle = \lambda_3 e^{-\left|t-t'\right|}\cos[\omega_{c}(t-t')]$.

Firstly, we apply the PTA in the computational basis to effectively enforce a Pauli channel and conduct experiments to acquire SpecTTM. 
In the upper panel of Fig.\ref{fig: nonPauli-RHP}, we compare $\Gamma_t$, estimated under different means, for this one-qubit system.
The red squares present that the rigorous $\Gamma_t$ that could be numerically determined without resorting to PTA for this simple model. The blue lines give $\Gamma_t$ after PTA in the computational basis.  In this case, we reconstruct $\Gamma_t$ (blue line) from SpecTTM dataset via Eq.~(\ref{eq:specttmb}) and Eq.~(\ref{eq:decoherencerate}). Clearly, the approximated result (blue lines) deviates significantly from the exact one (red squares).

Next, we consider the method discussed in Sec.\ref{sec:optimal-basis}] to search for an optimal twirling basis in order to better represent the correct dynamics with a Pauli channel. In the lower panel of Fig.\ref{fig: nonPauli-RHP}, we present $\Gamma_t$  (green lines) deduced from SpecTTM based on this optimal basis. A clearly shown, the approximated results under optimal basis agree well with the original one. In other words, the non-Markovianity of a non-Pauli channel can be maximally recovered with an optimal basis.

\section{Conclusion}

In this work, we propose SpecTTM for characterzing SPAM error-free spectrum of non-Markovian Pauli channels.  Our proposed protocol relies on SpecQPT to extract spectral information of a dynamical map, then the standard TTM relations can be alternatively cast into spectral versions for Pauli channels. As illustrated in this work, the SpecTTM approach can accurately capture every aspect of a non-Markovian Pauli channel. With access to the SpecTTM data,  it is possible to (1) assess the non-Markovianity, and (2) reconstruct the noise power spectrum. As argued earlier, Pauli channels (and generalized cases including amplitude damping channels with a diagonal R sub-matrix inside the Pauli transfer matrix) represent most popular noise and error models for quantum circuits due to various physical motivations.  Beyond Pauli channels, we also propose to identify an optimal basis to apply Pauli twirling approximation and collect the SpecTTM data in order to better characterize the dynamical process of interests.

While SpecQPT under PTA only captures partial information of a general quantum channel, the reconstructed SpecTTM still encodes useful information, such as giving an estimate on the degrees of non-Markovianity of a noisy process.  As we continuously improve quality of quantum devices, it becomes more crucial to have a simple approach to quantify non-Markovianity in a high-precision manner, for instance, complete removal of SPAM errors. Probing this subtle aspect of a noisy process will certainly help to design better quantum hardwares.

We acknowledge that SpecTTM is not a scalable protocol like randomized benchmarking.  The non-scalability certainly limits the size of a quantum device that could be subjected to such an analysis. Nevertheless, we note that SpecTTM is not necessarily consuming more resources (to conduct the data-acquisition experiments) than standard QPT.  Yet, one gets SPAM error-free estimates on the spectrum of dynamical processes, which usually requires a SPAM-resistant tomographic technique that consumes even way more resources.  Hence, we argue that SpecTTM is currently the method of choice that balances the trade-off between resource consumption and level of characterization if one desires to probe non-Markovianity of a noisy process in a quantum circuit.  

Finally, SpecTTM could be further optimized by adopting more sophisticated methods to reduce the experimental costs of performing QPTs such as the compressed sensing and other approaches.  Being able to conduct careful analysis on clusters of a small number of neighboring qubits in a connectivity-limited hardware architecture is among the most promising applications of SpecTTM in the NISQ era.

\appendix

\section{Signal functions for non-Markovian channel}\label{app:signalfunction}
We propose a protocol to construct a dynamical map $(\Lambda_n)^k$, which is the k-th multiplicative of a non-Markovian Pauli channel $\Lambda_n$ at time $t=t_n$. The corresponding signal function at time $t_n$ reads,
\begin{equation}
\begin{aligned} g^\prime_{t_{n}}(k)=& \frac{1}{2^n} \sum_{\mu=1}^{N} \operatorname{Tr}\left[P_{\mu} \mathcal{N}_{m} \circ \Lambda_{n} \circ\left(M_{\mu} \circ \mathcal{N}_{m} \circ \Lambda_{n}\right)^{k-1}\right.\\ & \circ \mathcal{N}_{p}\left(P_{\mu}\right)] \end{aligned},
\end{equation}
where 
\begin{equation}\label{eq:mu}
M_{\mu}[\rho]=P_{+} \rho P_{+}+P_{-} \rho P_{-}.
\end{equation} 
A Pauli matrix can be decomposed as follows: $P_{\mu}=\sum_+\left|\mu_{+}\right\rangle\left\langle \mu_{+}|-\sum_-| \mu_{-}\right\rangle\left\langle \mu_{-}\right|=\sum_+P_{+}-\sum_-P_{-}$, where
$\left|\mu_{+}\right\rangle$ denote eigenvectors with a positive eigenvalue, and $\left|\mu_{-}\right\rangle$ denote eigenvectors with a negative eigenvalue. Because $\operatorname{Tr}[P_\mu \Lambda_n(\sum_+P_+)]+\operatorname{Tr}[P_\mu \Lambda_n(\sum_-P_-)]=0$, we deduce that $\operatorname{Tr}[P_\mu \Lambda_n(P_\mu)]=2\operatorname{Tr}[P_\mu \Lambda_n(\sum_+P_+)]$. Thus, SpecQPT can be attained by preparing $d/2$ initial states for each $P_\mu$.
\begin{equation}
\begin{aligned} g^\prime_{t_{n}}(k)
=& \frac{1}{2^{(n-1)}} \sum_{\mu=1}^{N} \operatorname{Tr}\left[P_{\mu} \mathcal{N}_{m} \circ \Lambda_{n} \circ\left(M_{\mu} \circ \mathcal{N}_{m} \circ \Lambda_{n}\right)^{k-1}\right.\\ 
& \circ \mathcal{N}_{p}\left(P_{+}\right)] \\
=&\sum_{\mu=1}^N R_{\mu\mu}(k).
\end{aligned} 
\end{equation}

For Pauli channel, the R matrix is clearly diagonal. The construction of $k$-multiplicative of a dynamical map is simply related to construction of $k$-th power of the diagonal matrix $R_{\mu\mu}(1)$. For simplicity, we outline the proof for the single-qubit case without considering SPAM errors.
In particular, it is to verify the following relation for $g_{t_n}(1)$,
\begin{equation}
\begin{aligned}
R_{\mu\mu}(1)&=\operatorname{Tr} [P_\mu  \Lambda_n(P_+)] \\
&=(W_1-W_2),
\end{aligned}
\end{equation}
where $\left\langle \mu_+\left|\Lambda_{n}\left(P_{+}\right)\right| \mu_+\right\rangle=W_1$ and $\left\langle \mu_-\left|\Lambda_{n}\left(P_{+}\right)\right| \mu_-\right\rangle=W_2$. To facilitate the following discussion, we also introduce $\left\langle \mu_+\left|\Lambda_{n}\left(P_{-}\right)\right| \mu_+\right\rangle=W_3$ and $\left\langle \mu_-\left|\Lambda_{n}\left(P_{-}\right)\right| \mu_-\right\rangle=W_4$. Recall the definition of $M_\mu$ in Eq.~(\ref{eq:mu}), then it is also straightforward to establish the following relations for $g_{t_n}(2)$,
\begin{equation}
\begin{aligned}
R_{\mu\mu}(2)
&=\operatorname{Tr} [P_\mu  \Lambda_n \circ M_\mu \circ \Lambda_{n} (P_+)] \\
&=W_1\times W_1+W_3\times W_2-W_2\times W_1-W_4 \times W_2.
\end{aligned}
\end{equation}
Again, since $\operatorname{Tr}[P_\mu \Lambda_n(P_+)]+\operatorname{Tr}[P_\mu \Lambda_n(P_-)]=0$, we have $W_1-W_2=W_4-W_3$ and
\begin{equation}
\begin{aligned}
R_{\mu\mu}(2)
&=W_1\times W_1+W_2\times W_2-W_2\times W_1-W_1 \times W_2 \\
&=R_{\mu\mu}(1)^2
\end{aligned}
\end{equation}
If we repeat the analysis for the elements of $g_{t_n}(k)$ for $k=1,2,...,K$, then we should attain
\begin{equation}
R_{\mu\mu}(k)=R_{\mu\mu}(1)^k.
\end{equation}
So far, we discuss how the construction of $k$-multi;icative of a dynamical map can be realized by projective measurements. However, as this protocol relies on making $k$ measurements, the results also are also affected by the accompanied measurement errors,
\begin{equation}   
\begin{aligned}
g^{\prime}_{t_n}(k)&=\frac{1}{2^n} \sum_{\mu=1}^{N} \operatorname{Tr}\left[P_{\mu}\left(\mathcal{N}_{m} \circ \Lambda_{n}\right)^{k} \circ \mathcal{N}_{p}\left(P_{\mu}\right)\right] \\
&=\operatorname{Tr}\left[(R_{\text {meas }} R)^{k} R_{\text {prep }}\right] \\
&=\operatorname{Tr}\left[A_{\mathrm{P}} D^{k}\right]=\sum_{j=1}^{N} A_{j} (m_j\lambda_{j})^{k}
\end{aligned}
\end{equation}

In order to remove these measurement errors, we propose a supplementary protocol in which the original single projective measurement is replaced by double measurements. Going through the same analysis laid out for the single-measurement protocol described above, the double-measurement protocol leads to the following signal function,
\begin{equation} 
\begin{aligned}
g_{t_n}^{\prime\prime}(k)&=\frac{1}{2^n} \sum_{\mu=1}^{N} \operatorname{Tr}\left[P_{\mu}\left(\mathcal{N}_{m} \circ \mathcal{N}_{m} \circ \Lambda_{t i}\right)^{k} \circ \mathcal{N}_{p}\left(P_{\mu}\right)\right]   \\
&=\operatorname{Tr}\left[(R_{\text {meas }} R_{\text {meas }} R)^{k} R_{\text {prep }}\right]  \\
&=\operatorname{Tr}\left[A_{\mathrm{P}} D^{k}\right]=\sum_{j=1}^{N} A_{j} (m_j^2\lambda_{j})^{k}.
\end{aligned}
\end{equation}
As shown in Eq.~(\ref{eq:lambda}) of the main text, one can easily recover SPAM-error free eigenvalues of the Pauli channel by extracting the spectral information from $g_{t_n}^{\prime}(k)$ and $g_{t_n}^{\prime\prime}(k)$.

\bibliographystyle{IEEEtran}
\bibliography{main}

\end{document}